# Sustained biexciton emission in colloidal quantum wells assisted by dopant-host interaction

*Junhong Yu[+1], Manoj Sharma[+1], Mingjie Li[+2], Pedro Ludwig Hernández-Martínez[1], Savas Delikanli[1], Ashma Sharma[1], Yemliha Altintas[3], Chathuranga Hettiarachchi[5], TzeChien Sum, Hilmi Volkan Demir[*1,2,3], Cuong Dang[*1,4,5]*

[1]LUMINOUS! Centre of Excellence for Semiconductor Lighting and Displays, School of Electrical and Electronic Engineering, Nanyang Technological University, 50 Nanyang Avenue, 639798, Singapore

[2]School of Physical and Mathematical Sciences, Nanyang Technological University, 639798, Singapore

[3]Department of Electrical and Electronics Engineering and Department of Physics, UNAM-Institute of Materials Science and Nanotechnology, Bilkent University, Bilkent, Ankara, 06800, Turkey

[4]CINTRA UMI CNRS/NTU/THALES 3288, Research Techno Plaza, 50 Nanyang Drive, Border X Block, Level 6, 637553, Singapore

[5]Centre for OptoElectronics and Biophotonics, School of Electrical and Electronic Engineering & The Photonics Institute, Nanyang Technological University, 50 Nanyang Avenue, 639798, Singapore

[+] Junhong Yu, Manoj Sharma and Mingjie Li contribute equally to this work

* Email: volkan@stanfordalumni.org; hcdang@ntu.edu.sg

**Abstract:** Biexcitons have been considered as one of the fundamental building blocks for quantum technology because of its overwhelming advantages in generating entangled photon pairs. Although many-body complexes have been demonstrated recently in mono-layer transition metal dichalcogenides (TMDs), the low emission efficiency and scale up capability hinder their applications. Colloidal nanomaterials, with high quantum efficiency and ease of synthesis/processing, are regarded to be an appealing complement to TMDs for biexciton sources. However, a progress towards biexciton emission in colloidal nanomaterials has been challenging largely by small binding energy and ultrafast non-radiative multiexciton recombination. Here, we demonstrate room-temperature biexciton emission in Cu-doped CdSe colloidal quantum wells (CQWs) under continuous-wave excitation with intensity as low as ~10 W/cm$^2$. The characteristics of radiative biexciton states are investigated by their super linear emission with respect to excitation power, thermal stability and transient photophysics. The interaction between the quantum confined host carriers and the dopant ions increases biexciton binding energy by two folds compared to the undoped CQWs. Such strong binding energy together with suppressed Auger recombination and efficient, spectrally narrow photoluminescence in a quasi-2D semiconductor enables sustained biexciton emission at room temperature, providing a potential solution for efficient, scalable and stand-alone quantum devices.

Entangled photon pairs are natural candidates for basic entities ('qubits') in quantum information and quantum computation[1-3]. Commonly, parametric down conversion (PDC) in nonlinear optics is utilized to create entangled photon pairs[4]. However, PDC mechanism suffers from non-guaranteed single photon pair as well as extremely low non-linear efficiency[5]. Utilizing biexciton states in quantum emitters is a promising strategy to circumvent the limitations since the radiative decay of a biexciton has two channels: either emit a horizontally (*H*) polarized photon or a vertically (*V*) polarized photon[1-3,6]. If fine-structure splitting (FSS) is negligible, the generated two photons are indistinguishable in any other properties except the polarization: $\sqrt{2}|\phi\rangle = |H\rangle|H\rangle + |V\rangle|V\rangle$ [7]. Benefiting from the strong many-body interaction in atomically thin transition metal dichalcogenides (TMDs), the first observation of biexciton emission at 10 K using femtosecond pulse excitation was achieved 3 years ago[8]. Since then, lots of investigations have been conducted to demonstrate continuous-wave pumped biexciton states at room temperature in TMDs[7,9-14]. However, the synthesis (i.e. mechanical exfoliation, chemical exfoliation, and chemical vapor deposition) of high quality TMDs is still suffering from various issues[15-17] such as synthesizing cost, upscaling capability (curbed by flake size and film uniformity), emission efficiency (quantum yield: < 5%) or limited spectral tunability (constrained by the valley band). The use of colloidal nanomaterials offers an avenue for fully solution-processed, highly efficient, bottom-up biexciton sources, which in turn enable the easy integration of quantum emitters into a wide variety of optoelectronic quantum architectures[18,19].

Till now, the demonstration of biexciton states in colloidal nanomaterials can only be achieved under intense excitation with ultrashort pulsed lasers and cryogenic condition due to the two key challenges: very fast Auger decay[20,21], and small biexciton binding energy[22,23]. The former relates to enhanced non-radiative recombination of multi-exciton states in nanometer confinement with a typical time constant of tens to hundred picoseconds, which leads to ultrafast deactivation of the high-order correlated excitonic states. The latter relates to Coulomb interaction to form a bounded four-body complex from two excitons. The biexciton binding energy in colloidal nanomaterials[22] is typically smaller than 30 meV, comparable to the thermal energy at room temperature and much smaller than the inhomogeneous broadening of exciton emission. Insufficient binding energy causes the thermal dissociation of biexciton states at room temperature and makes biexcitonic transition inseparable from exciton states. Suppression of Auger recombination has been investigated extensively for colloidal nanocrystals to achieve optical gain practically. Large-volume nanocrystals leveraging the well-known 'volume scaling'[21,24] or interfacial alloying core/shell nanocrystals[20,25] utilizing the smoothed confinement potential exhibit relatively long Auger recombination (nanosecond timescales). Unfortunately, large-size nanomaterials which lose the benefits of strong quantum confinement (i.e. large binding energy and reduced dielectric screening)[24,26] and smooth interface nanocrystals which associate with broad photoluminescence (PL) spectrum (FWHM: > 30 nm, caused by the interface composition gradient)[20,27] are hard to be ideal biexciton sources.

The requirements for colloidal nanomaterials which reconcile suppressed Auger recombination, quantum confinement and narrow emission spectrum can be satisfied with colloidal quantum wells (CQWs)[18,28,29]. In this quasi-2D system, the exciton wave functions are quantized only in vertical direction, which makes Auger process hard to meet the momentum conservation compared with the situation in colloidal quantum dots[18,22]. Also, the exciton emission in CQWs exhibits extremely narrow linewidth at room temperature (FWHM of 4-monolayer CdSe CQWs: ~40 meV) due to vertical thickness control at the single atomic layer level and weak phonon coupling[30-34]. However, the highest biexciton binding energy reported for CQWs (~ 34 meV) is still smaller than their emission bandwidth, and thus the biexciton emission peak is only resolved at cryogenic temperature[22,31,33].

Brovelli and colleagues[35] have shown that in Ag doped CdSe colloidal quantum dots (CQDs), by exploiting the effect of excitonic carriers in host nanomaterials on dopant ions, the nonmagnetic dopants can exhibit optically switchable magnetism. $Ag^+$ dopants are not magnetic ($[Kr]4d^{10}$, there is no unpaired electron spin in full 4d shell), after capturing the photo-generated hole in host CdSe, $Ag^+$ is oxidized to $Ag^{2+}$ ($[Kr]4d^9$) and the unpaired electron spin can exhibit paramagnetic behavior. Reversibly, we propose that in doped CdSe CQWs, through utilizing the effect of dopant ions on the excitonic carriers in host nanomaterials, the biexciton binding energy can be enhanced. The dopants strongly capturing the photo-generated holes dilute the carrier density in the host nanomaterial, thus increases Debye screening length (reduces the dielectric screening) leading to increased biexciton binding energy[36-38]. In this work, we have demonstrated Cu dopants in CQWs enhances the biexciton binding energy up to 64 meV, a factor of 2 greater than the values reported in undoped counterparts. The boosted biexciton binding energy, suppressed Auger, and narrow emission spectrum enable continuous-wave biexciton emission at room temperature. A comprehensive analysis of this sustainable biexciton emission in Cu-doped CdSe CQWs (Cu:CdSe CQWs) is presented via the carrier dynamics (time-resolved photoluminescence and transient absorption) with temperature and excitation dependent emission spectra.

### Cu:CdSe CQWs as a dopant-host system

We prepared Cu-doped CdSe CQWs according to the existing recipe[39] and optimized the procedures to achieve highly efficient and reproducible synthesis (details of the synthesis and material characterization are described in Supplementary Note 1). The structure information of Cu:CdSe CQWs is shown in Supplementary Fig. S1: the four-monolayer (4 ML) CQWs are 1.2 nm thick[18] and tens of nanometers along each side laterally. The PL quantum yield (PLQY) of Cu:CdSe CQWs is in the range of 40% to 80% depends on the amount of Copper precursor added in the synthesis (Fig. S2). In terms of biexcitonic emission behavior, CQW with ~30 Cu dopants (see Methods) is tested in this work unless otherwise mentioned.

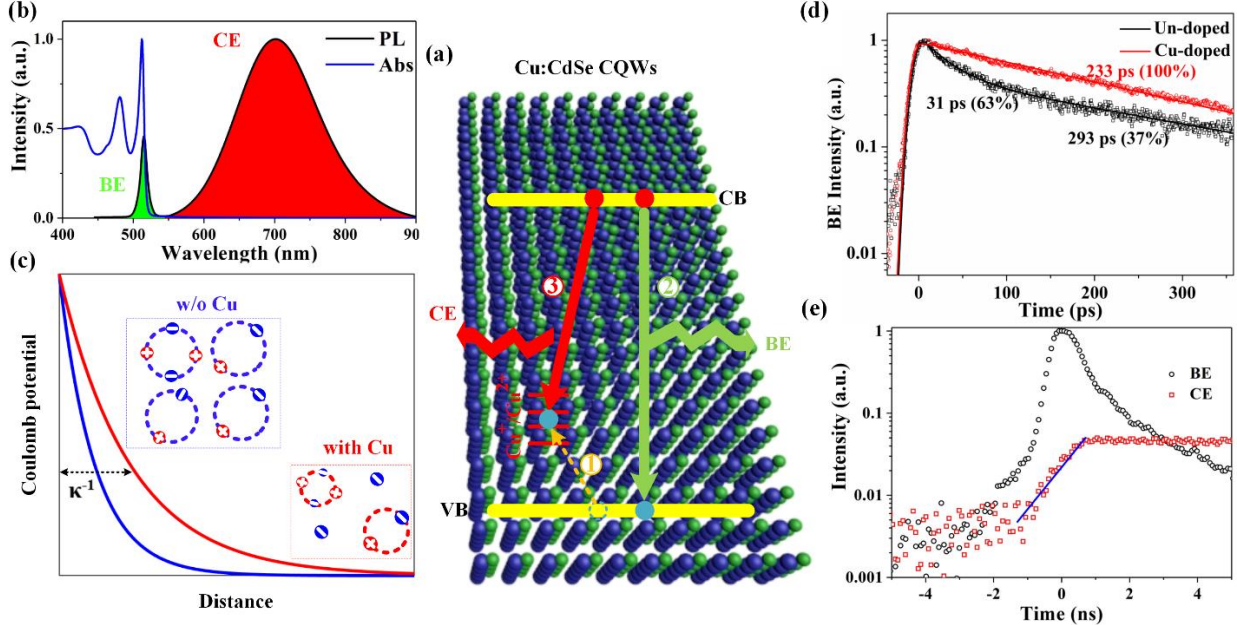

**Figure 1. The dopant-host system in a Cu:CdSe CQW. CE: copper-related emission; BE: band edge emission. (a)** Schematic of carrier dynamics in the Cu-doped CQWs with photo excitation. After capturing of a VB hole (process 1), $Cu^+$ is changed to $Cu^{2+}$, activating the CE through recombining with a CB electron (process 3). **(b)** Optical absorption and steady-state PL spectrum of Cu-doped CQWs in solution under Xeon lamp illumination, green color region indicates the BE and red color region indicates the CE. **(c)** Conceptual illustration of the reduced Coulomb screening in Cu-doped CQWs. Utilizing the hole-capture ability of Cu dopants, the carrier density in host CdSe CQWs has been diluted and longer Debye length can be expected. Red curve: with copper dopants. Blue curve: without copper dopants. **(d)** Normalized decay curves of BE for undoped CQWs and Cu-doped CQWs under the pump fluence of 0.4 μJ/cm$^2$. The PL dynamics of both doped and undoped CQWs are probed at 514 nm. The difference in dynamics sheds light on the suppressed intrinsic traps and the hole capture rate of Cu-dopants. **(e)** Time-resolved photoluminescence of BE and CE in Cu-doped CQWs with the pump fluence of 0.4 μJ/cm$^2$. The black circles show the excitonic dynamics (514 nm) and the red squares presents the copper ML$_{CB}$CT dynamics (integrated between 650 and 750 nm). Fitting the CE dynamics (blue solid line) yields the time constant of hole capture process (~1 ns), which is much slower compared with the Cu-doped quantum dots[19].

The copper dopants in CQW host behave like a hole-trapper[40-44], as illustrated in Fig. 1a. After excitation, copper dopants irreversibly localize the photo-generated holes in CQW host, $Cu^+$ is promoted to $Cu^{2+}$ and therefore activated as a radiative acceptor for the conduction band (CB) electrons, which can be expressed as[35]: $[Ar]3d^9 + e \rightarrow [Ar]3d^{10} + h\nu$ ($h\nu$ is the photon energy of Cu-related emission and e denotes an electron in CB)[19,39]. As a result, the photoluminescence (PL) spectrum (Fig. 1b) of Cu:CdSe CQWs is composed of two emission bands: (1) a narrow band-edge emission (BE) at ~514 nm with a full-width at half-maximum (FWHM) of ~40 meV, which is not affected by copper dopants; (2) a broad Cu-related emission (CE) peaking at ~700 nm with a very large FWHM of ~350 meV, the broad emission linewidth is arising from a wide distribution of copper energy level rather than size/doping inhomogeneity within an ensemble of Cu:CdSe CQWs, as detailed previously[40,41,44,45]. The absorption spectrum (Fig. 1b) of Cu:CdSe CQWs is also consistent with the scenario presented in Fig. 1a: two clear excitonic features associated with electron to heavy-hole (512 nm) and electron to light-hole (480 nm) transitions, remain unchanged compared with undoped CdSe CQWs (absorption spectra of undoped and doped CQWs with different copper doping levels are presented in Fig. S2). Notably, there exists a weak and broad absorption tail on the red side of the first CQW excitonic maximum (Fig. S2), which can be assigned as the direct transition of d orbital electrons in $Cu^+$ to the CB of CdSe CQWs[39,41,43]. However, the contribution of this transition to BE and CE is eliminated since the extinction coefficient of this tail at its maximum is smaller than 1000 M$_{Cu}^{-1}$cm$^{-1}$ (Fig. S3a) and without the hole-capture process, the extremely low recombination rate (0.002 ns$^{-1}$, Fig. S3b) between $Cu^{2+}$ and CB electron cannot compete with the band-edge recombination.

In quantum-confined nanomaterials, the screened Coulomb potential for electrons and holes can be reduced with longer Debye screening length (in neutral systems, $L_{Debye} \propto 1/\sqrt{<N>}$, where $<N>$ is the number of photo-generated excitons per nanoparticles)[46-47]. Several groups have demonstrated that the binding energy in nanomaterials can be significantly weakened via free electron injection since the screening length ($L_{Debye} \propto 1/\sqrt{<N>+n_e}$, where $n_e$ is the number of injected electrons) is smaller compared to the neutral case[36,37]. Accordingly, in Cu:CdSe CQWs, as conceptually illustrated in Fig. 1c, the hole-localization ability of $Cu^+$ can strongly dilute the hole density and expand the screening length in host CdSe CQWs ($L_{Debye} \propto 1/\sqrt{<N>-\alpha n_{Cu}}$, where $\alpha$ is a ratio to consider the widely distributed energy levels of $Cu^+$). Thus, enhanced many-body interaction can be expected in this host-dopant system.

In view of this host-dopant system, the diluted hole concentration in CQW host is suggested by strong CE (Fig. 1b). It is important to ensure the hole-capture dynamics can support the radiative many-body states since the ultrafast hole consumption leads to insufficient constituent (holes) for many-body complexes. To figure out the kinetics of hole capturing in Cu:CdSe CQWs, Fig. 1d presents the BE dynamics of undoped and doped CQWs in the sub-nanosecond time scale, which shows that the ultrafast recombination channel (~31 ps) is absent in doped CQWs. The single exponential decay behavior of Cu:CdSe CQWs in a short time window is consistent with the theoretical calculation conducted by Nelson[44,48], in which highly localized covalent [CuSe$_4$] cluster can suppress or deactivate intrinsic traps and thus, enhance the quantum efficiency of Cu-doped nanocrystals (Fig. S2). Here, we can assume that Auger can be negligible in low pumping regime ($<N>$ is smaller than 0.1), only hole capture process is responsible for the difference in BE dynamics between undoped and doped CQWs. Thus, we can extract the time constant of hole capture process[49] ($\tau_{hc}$) from exciton lifetime of doped ($\tau_{doped}$) and undoped ($\tau_{undoped}$) CQWs using $1/\tau_{undoped} + 1/\tau_{hc} = 1/\tau_{doped}$, which is around 1.1 ns.

Furthermore, since the prerequisite of CE is $Cu^+$ localizes a hole, hole-capturing time can be calculated from the rising part of CE[43]. As shown in Fig. 1e, in a nanosecond time window, a slow rising process in CE indicating a time constant of 977±45 ps is in good agreement with hole-capture dynamics extracted from Fig. 1d. It is worth to mention that, the onset of CE is significantly delayed compared to BE, emphasizing again the proposed carrier dynamics in Fig. 1a. In addition, as shown in Fig. S4, by comparing the gated emission spectra of Cu:CdSe CQWs when CE is absent (integrated time window: 0~1 ns), and when both CE and BE are present (integrated time window: 1~5 ns), we can draw the same conclusion that the time constant of hole-capture process in Cu:CdSe CQWs is prolonged and in the range of nanosecond. Compared with the ultrafast hole-capture process (an averaged lifetime of ~25 ps) in Cu:CdSe CQDs[43], we attribute the slow dynamics in CQWs to the larger spatial separation of band-edge excitons and copper dopants due to large lateral size of the structure. The 2D morphology can slow down the hole localization, suppress the Auger recombination and lengthen the lifetime of multiple-exciton, as four-exciton states observed in CdSe/CdTe CQWs using transient absorption spectroscopy[50,51].

**Continuous-wave excited biexcitons**

We have therefore obtained a colloidal nanomaterial with several unique properties that are crucial to radiative many-body complexes: reduced Coulomb screening, slow hole localization and suppressed Auger. To identify the potential many-body complexes in Cu:CdSe CQWs, PL of the drop-cast film under continuous-wave excitation with different intensity at room temperature has been studied (Fig. 2a and Fig. 2b). With low pump intensity (< 6 W/cm$^2$, the bottom spectrum in Fig. 2b), the BE shows single symmetric peak located at ~514 nm, which is assigned to typical single exciton emission[4,5]. With gradually increased excitation intensity, a new emission feature appears at the low-energy side (peaking at ~528 nm) and the intensity of this new feature grows much faster than the intensity of exciton emission with increasing excitation intensity (see the recorded movie in Supplementary Materials). Beyond the excitation intensity of ~10 W/cm$^2$, the new feature emerges as the strongest emission channel (see Fig. 2c, the maximum intensity of BE shifts from 514 to 528 nm with higher excitation intensity).

To quantitatively analyze the properties of the emerging emission feature, BE is fitted by two separately symmetric emissions using Voigt function[22,52], as shown in Fig. 2b. The fitting profiles are in good agreement with the experimental spectra with different excitation intensities. After spectrally separating the exciton emission and the new emission feature, Fig. 2d presents the integrated emission intensities as a function of the excitation intensity. The excitation intensity dependence of these two features can be described adequately by the power-law equation[8-12]: $I_{emission} \propto I_{Intensity}^k$. The exciton emission exhibits a slightly sublinear fluence dependence ($k = 0.91$), due to the hole capturing process from host CQWs to Cu+ sites[8,43]. However, the integrated intensity of the new emerged emission displays a superlinear growth with excitation intensity ($k$ = 1.68). On the basis of nearly quadratic (our case is ~1.85) intensity dependence of the new emission feature with respect to the exciton emission, the new feature peaked at ~528 nm can only be assigned to biexcitons formed from two excitons, the population of each exciton growing almost linearly (~ 0.91) with excitation intensity[12]. This is similar to the case of biexcitons in other systems, providing a first evidence that Cu:CdSe CQWs support radiative biexcitons at room temperature with extraordinary low excitation intensity (~ 10 W/cm$^2$).

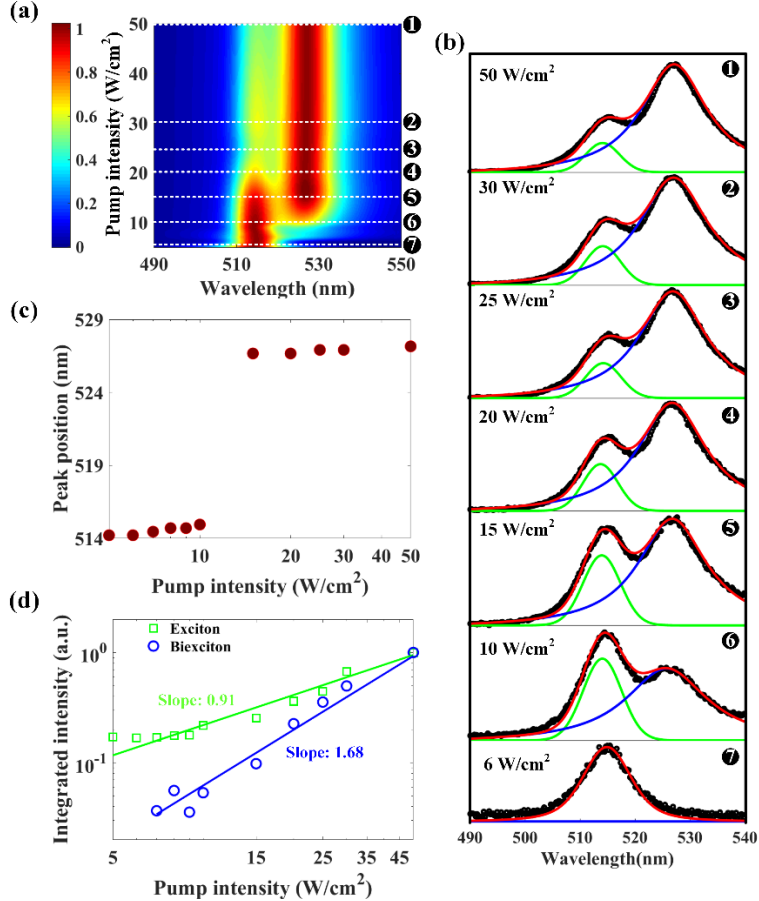

**Figure 2. Photoluminescence spectra of Cu-doped CQWs at room temperature with varied continuous-wave excitation intensities.** (a) Normalized BE map of Cu:CdSe CQWs under different excitation intensities. The white dashed lines indicate different excitation intensities and the ordered numbers correlate to the BE spectra in Fig. 2b. (b) BE spectra recorded for different excitation intensities. Black dots: measurement; green line: the fitting of exciton; blue line: the fitting of biexciton; red line: the overall fitting. (c) Wavelength of the maximum BE intensity as a function of the excitation intensity. Beyond ~10 W/cm$^2$, the intensity of biexciton emission (peaking at ~528 nm) exceeds the exciton emission (peaking at ~514 nm). (d) Logarithmic plot of the integrated emission intensity for exciton (red stars) and biexciton (blue circles) as a function of the excitation intensities. The green and blue solid lines are the power-law fitting for exciton and biexciton, respectively.

Furthermore, the biexciton binding energy can be extracted based on the energy difference between exciton and biexciton emissions assuming that the radiative decay of a biexciton will generate one photon and an exciton (see the derivation in Methods) [8,12]. The value of ~64 meV derived here is two-fold of the biexciton binding energy in undoped CdSe CQWs[22] (see PL spectra of undoped CdSe CQWs in Fig. S5). This is also higher than that of inorganic colloidal perovskites[52] and surprisingly comparable to the value in monolayer WSe$_2$[8]. Instead of classical explanation for the enhanced biexciton binding energy using reduced Coulomb screening as illustrated in Fig. 1c, we have performed quantum-mechanical calculation using a biexciton Hamiltonian with a perturbed electron-hole interaction potential by copper dopants to obtain deeper physical insight into the biexciton state. The simulation yields an enhancement of biexciton binding energy of ~30% in Cu:CdSe CQWs compared to undoped CdSe CQWs, which is in good agreement with the experimental result (see Fig. S11, Fig. S12 and details of the calculation in Supplementary Note 3).

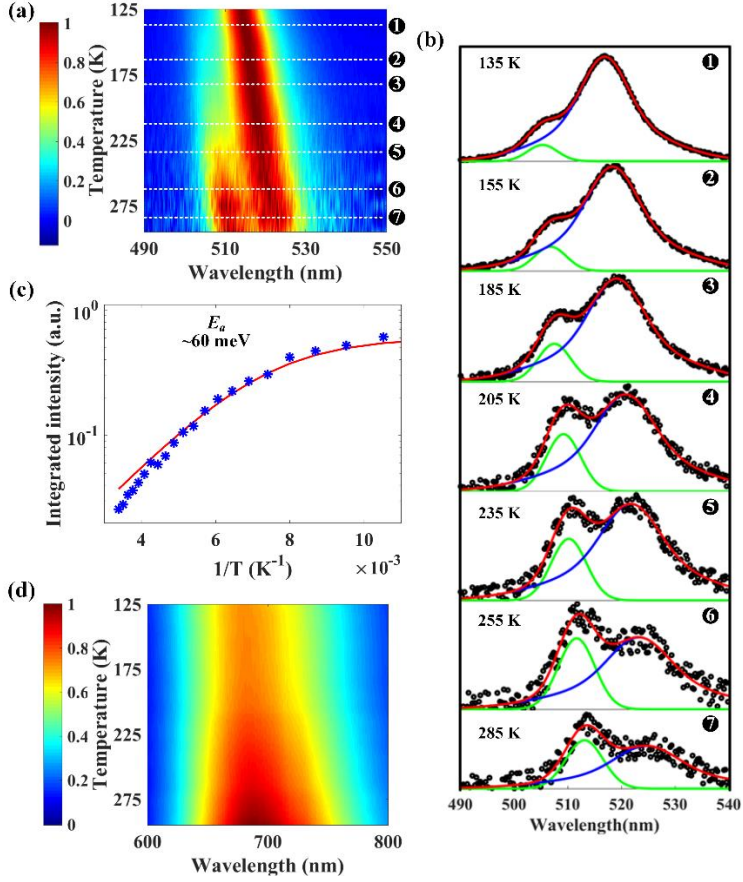

**Figure 3. Photoluminescence spectra of Cu-doped CQWs with fixed excitation intensity of 10 W/cm$^2$ at different temperatures.** (a) BE map of Cu:CdSe CQWs at different temperatures. The white dashed lines correspond to different temperatures and the ordered numbers correlate to the BE spectra in Fig. 3b. (b) BE spectra recorded at different temperatures. Black dots and red/green/blue lines have the same denotations as expressed in Fig. 2. (c) Integrated biexciton emission intensity (blue stars) as a function of 1/T. The red line is the Boltzmann distribution fitting, which yields an activation energy of ~60 meV. (d) Normalized CE map as a function of the temperature. the suppressed CE suggests the weaker hole capture ability of copper sites at low temperature.

To probe the biexciton states in more detail, we have investigated the thermal stability of biexcitons in Cu:CdSe CQWs at the excitation intensity of 10 W/cm$^2$. As temperature decreases from 298 to 125 K, besides the expected spectra change and increased emission intensity (Fig. 3a and Fig. 3b), biexciton emission displays a small intensity increasing rate when $T < 100$ K (Fig. 3c). This trend can be explained by the assumption[8] that the formation rate of biexciton is roughly constant with varied $T$, while the dissociation rate ($\gamma_{biex}$) is enhanced by the thermal energy: $\gamma_{biex} = \gamma_{const} + a*\exp(-E_a/k_BT)$, where $\gamma_{const}$ is a fixed recombination rate, $a$ is a ratio factor and $E_a$ is the activation energy. Meanwhile, the integrated emission intensity of biexciton could be well fitted by the Boltzmann distribution (see Methods)[6,8,53] with an activation energy $E_a$ of ~60 meV (Fig. 3c), for which $\exp(-E_a/k_BT)$ will have a negligible effect on $\gamma_{biex}$ and the integrated intensity of biexciton emission will show a plateau when $T$ is lower than 100 K. The $E_a$ value of ~60 meV is almost equal to the biexciton binding energy derived from steady-state PL analysis (Fig. 2). This agreement emphasizes again the peak located at ~528 nm arises from biexcitons, in which the dissociation rate can be tuned by the thermal energy. Moreover, with decreased temperature, CE exhibits barely blue-shifted and narrowing spectra, companying with decreased emission intensity (Fig. 3d). This trend is another supporting evidence of the spatial separation of dopant ions and host carriers (see the discussion in Fig. 1d and Fig. 1e): with gradually decreased temperature, thermally assisted carrier movement will be suppressed, weakening the hole capture probability and resulting in reduced CE intensity[41].

### Transient photophysics of biexcitons

To furtherly justify our assignment of biexcitons, we have investigated the transient photophysics in Cu:CdSe CQWs. Firstly, we conducted time-resolved PL measurement in Cu:CdSe CQW film with varying excitation fluences (see details in Methods). In the bottom panel of Fig. 4a, a narrow PL spectrum is observed with low excitation fluence (0.4 µJ/cm$^2$, the bottom panel); while with high fluence (0.4 µJ/cm$^2$, the top panel), biexciton emission appears and locates exactly at the wavelength observed in continuous-wave excited PL (Fig. 2 and Fig. 3). Following the white and black dashed lines in Fig. 4a, we compare the PL decay trace for X (located at 514 nm) and XX (located at 528 nm). With low fluence in which biexciton emission is absent, the dynamics of X and XX are almost identical (single-decay process with similar lifetime of ~234 ps and ~245 ps for X and XX, respectively), and their onsets are at the same time (see Fig. 4b). In contrast, with high fluence and thus high exciton density (Fig. 4b), the onset of XX is significantly delayed compared with X, accompanying a fast recombination process with the lifetime of ~128 ps while the lifetime of X is ~201 ps. This delayed and accelerated decay provides another strong evidence of the biexciton emission[6,54] at XX. Note that the lifetime ratio between exciton and biexciton is ~1.57, which is close to the value reported in traditional epitaxial quantum wells (GaAs or CdS)[55,56]. The result strongly suggests a suppressed Auger recombination in Cu:CdSe CQWs compared to that in type-I CdSe CQDs[20,21], where biexciton lifetime is typically two order of magnitude shorter than exciton lifetime. Moreover, according to the statistics of carrier recombination, the radiative recombination rate of the biexciton should be one quarter of the value in exciton, here, our ratio is smaller than 4 (ref. 21), excluding the efficient Auger recombination in Cu:CdSe CQWs, as discussed in previous studies[22,24,25].

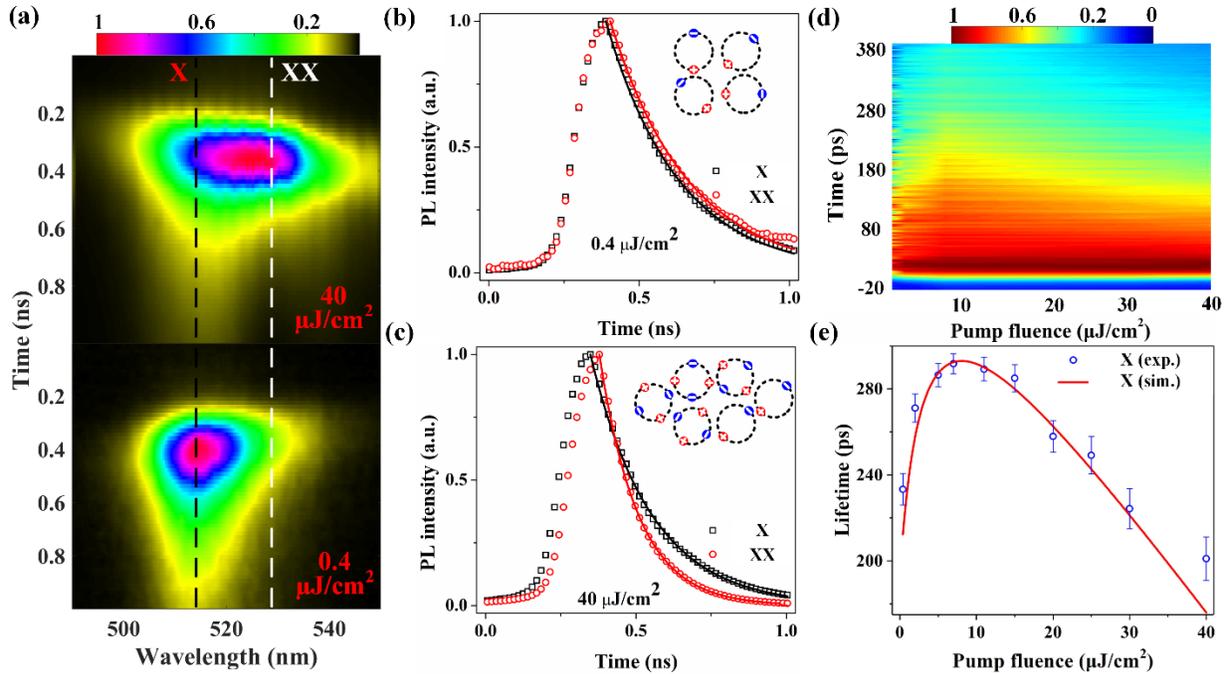

**Figure 4. Time-resolved photoluminescence of Cu-doped CQWs.** (a) Streak camera images of Cu:CdSe CQWs under excitation of 100 fs pulsed laser. Top panel: at high excitation fluence (40 µJ/cm$^2$) in which biexciton emission is observed. Bottom panel: at low excitation fluence (0.4 µJ/cm$^2$) in which biexciton emission is absent. Black dashed line is located at 514 nm (X: exciton) and white dashed line is located at 528 nm (XX: biexciton). (b) Decay traces probed at 514 nm and 528 nm and corresponding fitting curves when the excitation fluence is 0.4 µJ/cm$^2$. 514 nm (measurement: black squares, fitting: black solid line), 528 nm (measurement: red circles, fitting: red solid line). (c) Decay traces probed at 514 nm and 528 nm and corresponding fitting curves when the excitation fluence is 40 µJ/cm$^2$. Same color and shape coding as in (b). (d) Normalized PL intensity decay map of BE in Cu:CdSe CQWs (probed at 514 nm) as a function of the time and the excitation fluence. With low excitation fluence, the lifetime of X increases due to the filling of copper sites. Further increasing the excitation fluence, multiple carriers dominate the recombination dynamics of X. (e) Excitation fluence dependent lifetimes probed at 514 nm which is extracted from the 2D contour map shown in Fig. 4d. The red solid line is the fitting based on the coupled kinetic-equation model (see Fig. S9).

To shed more light on the effect of copper dopants on band edge emission in host CQWs, we have analyzed the fluence dependent dynamics of exciton (probed at 514 nm). As shown in Fig. 4d, initially increasing fluence results in a slower dynamics and the exciton lifetime will be close to the value in undoped CQWs (293 ps, Fig, 1d), indicating the fully-filled or saturated copper sites. This trend validates the proposed carrier mechanism in Fig. 1a and our method to extract the hole capture lifetime from host CQW to the Cu$^+$ sites (Fig. 1d). With even higher excitation fluence, in addition to the Cu$^+$ saturation, multiple-exciton interaction process is occurring with high excitation fluence and accelerates the exciton dynamics (two-exciton interaction process to promote into biexciton)[21,57]. We furtherly verify our interpretation by reproducing the fluence dependent exciton dynamics using a copper-perturbed coupled kinetic-equations (see Fig. S8)[8,58]. Briefly, exciton dynamics is controlled by the interplay between dopant-related exciton decay rate and fluence dependent exciton-exciton interaction process while biexciton formation rate is related to the thermal equilibrium status (see the model in Supplementary Note 2 for details). The quality of the fitting (see Fig. 4e) confirms the validity of our model: with low fluence, exciton dynamics is changing following the filling of $Cu^+$ site; Whereas with high fluence, multiple carrier interaction takes control the decay channel. Moreover, we can also predict the biexciton dynamics and fluence dependent biexciton intensity based on the proposed model, which is consistent with the experimental results (Fig. S10).

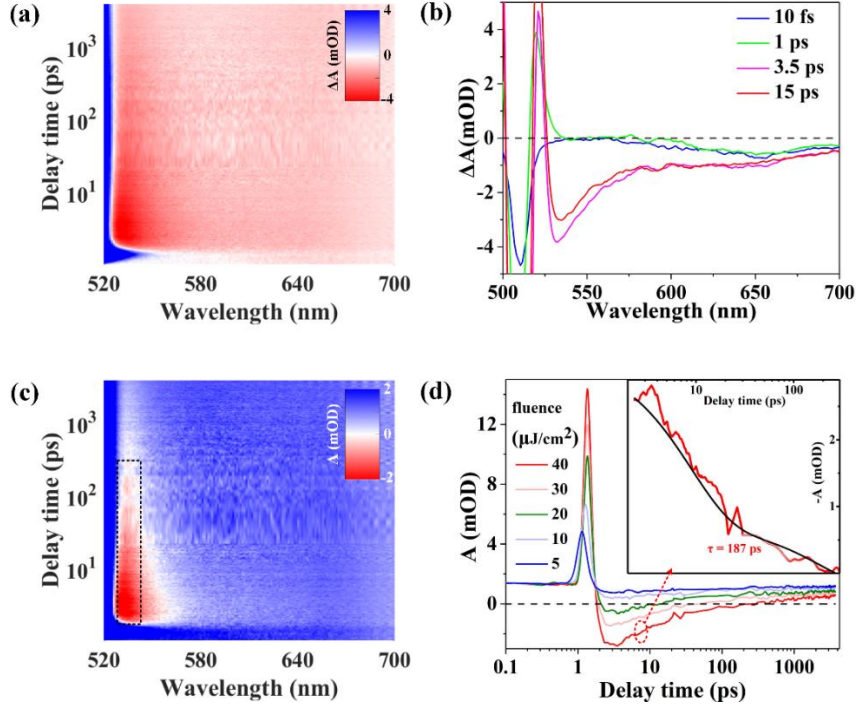

**Figure 5. Transient absorption spectra of Cu:CdSe CQWs, excitation laser: 100 fs @ 400 nm.** (**a**) 2D transient absorption change ($\Delta A$) images of Cu:CdSe CQWs (see the $\Delta A$ data with the full wavelength range in Fig. S6b). The measurement was obtained at a fluence of 40 μJ/cm². Also the negative $\Delta A$ beyond the PIA band is only observed in Cu:CdSe CQWs (see the $\Delta A$ images of undoped CQWs in Fig. S6a and S6d ). (**b**) $\Delta A$ spectra as a function of wavelength using different delay time at a fluence of 40 μJ/cm². The negative band around 528 nm exhibits a delayed formation. (**c**) 2D transient absorption ($A$) images of Cu:CdSe CQWs at a fluence of 40 μJ/cm². The red color-coded region indicates the stimulated emission ($A < 0$) around 528 nm. (**d**) Time-resolved $A$ spectra (probed at 528 nm) with different fluences. At a fluence of 40 μJ/cm², the gain can last over a period of >200 ps. The inset shows the dynamics of -$A$ at a fluence of 40 μJ/cm² which yields a lifetime of 187±5 ps.

A most striking observation is the stimulated emission in dispersions of Cu:CdSe CQWs at high excitation fluence in transient-absorption spectra (see Methods), arising from the radiative recombination of biexcitons. As expected, in the $\Delta A$ spectra of Cu:CdSe CQWs, we observe a weak absorption bleach (the transition of d orbital electrons in $Cu^+$ to the CB of CdSe CQWs, $ML_{CB}CT$, which is also discussed in Fig. 1) covering the whole spectrum beyond the red side of the positive photo-induced absorption (PIA) band, as shown in Fig 5a and 5b. This weak absorption bleach exhibits nearly invariant dynamics within the measured time window (> 3ns) which is consistent with the slow recombination between CB electrons and copper-trapped holes (lifetime: ~450 ns, Fig. S3). Notably, after several picoseconds delay, we unambiguously observe the appearance of a strong negative signal (the saturated red region in Fig. 5a) peaked at ~ 528 nm below the PIA band. Here, we can assign the emerged negative signal around 528 nm to a stimulated emission process, because of three reasons as follows. (**i**) This does not arise from $ML_{CB}CT$ since this process induced bleaching should occur simultaneously with the heavy-hole bleaching (see the curve in Fig. 5b with a delay time of 10 fs)[43]. Moreover, the strong negative signal exhibits much faster decay than the absorption bleaching of $ML_{CB}CT$, implying that the negative signal around 528 nm is related to the dynamics of band edge carriers[39]. (**ii**) The negative signal peaking at 528 nm shows a delay formation and only appears after several picosecond (Fig. 4b) with respect to the linear absorption bleaching band[58,59]. (**iii**) The negative signal can only be observed under high pump fluence (see the images of $\Delta A$ in Cu:CdSe CQWs with different pump fluences in Fig. S6e) and the energy location is exactly matching our biexciton emission spectra[50,58,59].

The stimulated emission from biexcitons is further verified by the absolute absorbance $A(\lambda,t) = \Delta A(\lambda,t)+A_0(\lambda)$, where $A_0(\lambda)$ is the linear absorption with the same measurement condition[59,60]. Fig. 5c presents the contour map of absolute absorbance at a pump fluence of 40 μJ/cm$^2$ which codes the positive absorbance with blue color and gain ($A<0$) with red color. The optical gain region is exactly matching the biexciton emission resolved in previous part (dashed black rectangle) and can last more than 100 ps. The dynamics of stimulated emission with different pump fluences probed at the wavelength of 528 nm is shown in Fig. 5d. with low fluence, no gain can be achieved. With increasing fluence (> 10 μJ/cm$^2$), the value of $A$ turns negative after the initial short-lived (< 1 ps) positive feature caused by hot carriers. Significant amplification is observed at a pump fluence of 40 μJ/cm$^2$ with a minimum absorbance value of ~2.4 mOD (ASE is also demonstrated in Cu:CdSe CQW film with a low threshold of ~53 μJ/cm$^2$, see Fig. S7). At the fluence of 40 μJ/cm$^2$, the optical gain relaxation is fitted to a bi-exponential decay. The slow decay component with a lifetime of ~187 ps matches well with the PL decay probed at 528 nm (~128 ps, Fig 4c), confirming again the long-lived biexciton states support the observation of sustainable biexciton emission in Cu:CdSe CQWs[22,50]. Interestingly, the gain dynamics shows a fast component (with a lifetime of 8.67 ± 0.2 ps) which is absent in the trPL measurement. This fast decay in absorbance is likely to be assigned to multi-exciton recombination.

## Discussion

We have discovered the four-particle neutral biexciton in Cu-doped CdSe CQWs, which is unambiguously identified by its transient photophysical properties, superlinear emission intensity, and thermal stability. Further studies are necessary to understand formation mechanism and constituent of biexciton, where the roles of a dark exciton was not observed in our Cu:CdSe CQWs. This is in contrast with the biexciton mechanism in WSe$_2$ where slow dark excitons enable biexciton states with continuous-wave excitation[12]. It is also worth to explore whether biexcitons in Cu-doped CQWs could have a more significant fine structure splitting since the energy difference between photon pair generated from biexciton can reduces the pure polarization entanglement substantially[13]. More evidences to characterize the electron-hole exchange interaction and optical selection rules could furtherly check the potential of biexciton in Cu-doped CQWs as practical entanglement photon sources.

Nevertheless, we have demonstrated biexciton emission in Cu-doped CdSe CQWs under continuous-wave excitation at room temperature. The achievement is attributed to the signatures of the material: large biexciton binding energy enhanced in host-dopant system and suppressed non-radiative Auger recombination mediated in 2D confinement structure. The observation of spectrally separable, radiative biexciton states is a key to study coherent many-body physics, such as condensation and superfluidity. Further, such sustainable biexciton emission at room temperature, using fully solution-processed colloidal assemblies of nanomaterials, offers broad potential applications in practical quantum optoelectronics, including quantum logic gates and high-order correlated photon emission.

## Methods

**Linear optical characterization of Cu:CdSe colloidal quantum wells.** Absorption spectra of Cu:CdSe CQWs in hexane is measured by a UV-VIS spectrophotometer (Shimadzu, UV-1800). PL spectra of Cu:CdSe CQWs in hexane are recorded using a fiber-coupled ANDOR spectrometer (monochromator: ANDOR Shamrock 303i, CCD: ANDOR iDus 401) with a diode laser excitation (Cobolt 06-MLD, excitation wavelength: 405 nm). Quantum yield (QY) of Cu:CdSe CQWs in hexane is measured with an integrating sphere and calculated as the ratio of absolute emission and absorption photons. The accuracy of the QY measurement is verified using Rhodamine 6G, whose QY of 94.3% measured in our setup, is found in good agreement with the standard value of 95%.

**Estimation the Cu atomic level.** With Inductively coupled plasma mass spectrometry (ICP:MS) we estimated Cu atomic levels with respect to cadmium and selenium. Average dimensions of our 4 ML Cu:CdSe CQWs measured by TEM microscopy are (36±1.9)×(15.0±1.5)×1.2 nm, which suggests 13500 cadmium and 10600 selenium atoms in one CQW. Therefore, using ICP:MS measurements we can estimate Cu atoms per CQW.

**Time-resolved photoluminescence measurement.** Time-resolved PL (TRPL) measurements are performed with a streak camera from Optronics. The 400-nm pump laser pulses for TRPL are generated from a 1000 Hz regenerative amplifier (Coherent LibraTM). The beam from the regenerative amplifier has a center wavelength at 800 nm, a pulse width of around 150 fs and is seeded by a mode-locked Ti-sapphire oscillator (Coherent Vitesse, 80MHz). 400-nm pump laser was obtained by frequency doubling the 800-nm fundamental regenerative amplifier output using a BBO crystal. All measurements are performed in the solid film at room temperature in ambient air (53±2% humidity) conditions.

**Continuous-wave pumping photoluminescence spectroscopy.** Continuous-wave excitation PL study is performed with a diode laser (Cobolt 06-MLD, excitation wavelength: 405 nm) and a fiber-coupled ANDOR spectrometer (monochromator: ANDOR Shamrock 303i, CCD: ANDOR iDus 401). Samples are drop-casted on a glass substrate for both room and low temperature measurements. The measurement is conducted in a surface emission geometry to avoid the spectra shift caused by reabsorption. The excitation spot radius is 50 μm produced by a plano-concave lens with a focal length of 75 mm. For the low temperature CW PL measurement, sample are cooled with a closed-cycle helium cryostat.

**Calculation of the biexciton binding energy based on emission spectra.** The biexciton binding energy is defined as the energy difference between biexciton states and two free excitons: $\Delta_{biex} = 2E_{ex} - E_{biex}$, where $\Delta_{biex}$ is the biexciton binding energy; $E_{ex}$ and $E_{biex}$ is the energy of exciton and biexciton, respectively. Considering that the radiative recombination of one biexciton will generate one exciton and emit one photon: $E_{biex} = E_{ex} + h\nu_{xx} = h\nu_x + h\nu_{xx}$ ($h\nu_x$ and $h\nu_{xx}$ is the photon energy of exciton and biexciton emission, respectively). Thus, we can calculate the biexciton binding energy ($\Delta_{biex}$) based on the spectra shift: $\Delta_{biex} = h\nu_x - h\nu_{xx}$.

**Thermal intensity distribution fitting.** As explained in Fig. 2g, assuming the formation rate of biexciton is roughly constant with varying temperature, while the decay rate ($\gamma_{biex}$) is enhanced by the thermal energy, the temperature dependent biexciton emission intensity change can be fitted by a Boltzmann distribution function: $b/[1+a*\exp(-E_a/k_B T)]$, where $b$ is a constant, $a$ is the coefficient, and $E_a$ is the activation energy which is corresponding to the biexciton binding energy.

**Transient absorption spectroscopy.** TA spectroscopy is performed using a Helios$^{TM}$ setup (Ultrafast Systems LLC) and in transmission mode with chirp-correction. The white light continuum probe beam (in the range of 400-800 nm) is generated from a 3 mm sapphire crystal using 800 nm pulse from the regenerative amplifier as mentioned in TRPL measurement. The pump beam spot size is ~0.5mm. The probe beam passing through the sample was collected using a detector for UV–Vis (CMOS sensor). All measurements are performed at room temperature in solution (hexane).

## Acknowledgements

We would like to acknowledge the financial support from Singapore National Research Foundation under the Program of NRF-NRFI2016-08, the Competitive Research Program NRF-CRP14-2014-03 and Singapore Ministry of Education AcRF Tier-1 grant (MOE-RG178/17). H.V.D is also grateful to acknowledge additional financial support from the Z4BA. We would like to thank Muhammad Taimoor and Thomas Kusserow at University of Kassel (Kassel, Germany) for carefully reading our manuscript.

## Author Contributions

C.D. and H.V.D supervised and contributed to all aspects of the research. J.Y., M.S, H.V.D and C.D. wrote the manuscript. J.Y. conducted the spectroscopy measurement and proposed the coupled kinetic model. M.S performed the material synthesis and designed them to achieve the best Biexciton lasing performance. S.D and A.S helped in material synthesis and characterizations. M.L. performed lifetime and transient absorption measurement. M.L. and TC.S. supervised the ultrafast dynamic analysis. P.L.H-M conducted the Hamiltonian calculation to predict the biexciton binding energy. Y.A. conducted the material characterization. All authors analysed the data, discussed the results, commented on the manuscript and participated in manuscript revision.

## Competing financial interests

The authors declare no competing financial interests.